\documentclass[12pt]{iopart}
\usepackage{graphicx}

\def\beq{\begin{eqnarray}}
\def\eeq{\end{eqnarray}}
\def\beqa{\begin{eqnarray}}
\def\eeqa{\end{eqnarray}}

\begin{document}

\title[Electron-phonon interaction dressed by electronic correlations]{Electron-phonon interaction dressed by electronic correlations near charge ordering as the origin for superconductivity in cobaltates}

\author{A Foussats$^1$, A Greco$^1$, M Bejas$^{1,2}$ and A Muramatsu$^2$}
\address{
$^1$ Facultad de Ciencias Exactas, Ingenier\'{\i}a y Agrimensura and
Instituto de F\'{\i}sica Rosario
(UNR-CONICET), Av. Pellegrini 250-2000 Rosario-Argentina
}
\address{$^2$ Institut f\"ur Theoretische Physik III, Universit\"at Stuttgart,
Pfaffenwaldring 57, D-70550 Stuttgart, Germany}

\eads{\mailto{foussats@ifir.edu.ar}, \mailto{agreco@fceia.unr.edu.ar}}

\begin{abstract}
We consider possible routes to superconductivity in hydrated
cobaltates $Na_xCoO_2.yH_2O$ on the basis of the $t$-$J$-$V$ model
plus phonons on the triangular lattice. We studied the stability
conditions for the homogeneous Fermi liquid (HFL) phase against
different broken symmetry phases. Besides the $\sqrt{3}\times
\sqrt{3}$-CDW phase, triggered by the nearest-neighbour Coulomb
interaction $V$, we have found that the HFL is unstable, at very
low doping, against a bond-ordered phase due to $J$. We also
discuss the occurrence of phase separation at low doping and $V$.
The interplay between the electron-phonon interaction and
correlations near the $\sqrt{3}\times \sqrt{3}$-CDW leads to
superconductivity in the unconventional next-nearest neighbour
$f$-wave (NNN-$f$) channel with a dome shape for $T_c$ around $x
\sim 0.35$, and with values of a few Kelvin as seen in
experiments. Near the bond-ordered phase at low doping we found
tendencies to superconductivity with $d$-wave symmetry for
finite $J$ and $x<0.15$. Contact with experiments is given along the paper.
\end{abstract}



\maketitle

\section{ Introduction}

Since superconductivity was discovery in hydrated cobaltates
$Na_xCoO_2.yH_2O$ for $x \sim 0.35$ and $y \sim 1.3$
\cite{takada03} an enormous amount of attention was focused on
this system, in spite of the rather modest critical temperature
$T_c$, that follows a characteristic dome shape \cite{schaak03}
with maximum value $T_c \sim 5K$ around a doping $x \sim 0.35$.
Indeed, since cobaltates may be considered as electron-doped
Mott insulators with a layered structure, where $Co$ ions are in a
low spin state ($S=1/2$) on a triangular lattice, it was expected
that the resonating valence bond (RVB) scenario proposed long ago
by Anderson \cite{anderson87} for the cuprates could be clearly
realized in the cobaltates.

The importance of strong electronic correlations, already advanced
in view of the small bandwidth of the ($t_{2g}$) levels close to
the Fermi energy \cite{singh00,singh03}, was confirmed by recent
photoemission studies \cite{hasan04,yang04} showing for $x=0.3$ a
reduction of bandwidth by a factor two with respect to the
calculated ones \cite{singh03}. In fact, even taking the whole
$t_{2g}$ manifold, the bare bandwidth is $\sim 1.6$ eV
\cite{singh00}, while from core level photoemission spectroscopy
\cite{chainani04}, values of the onsite interaction $U_{dd} \sim
3.0 - 5.5$ eV were estimated. Hence, electronic correlations seem
to play an essential role, so that Hubbard or $t$-$J$ models were
proposed \cite{baskaran03} to describe cobaltates.

Several unconventional pairing channels and mechanisms were
proposed for the superconducting state:
a) pure $t$-$J$ model RVB based calculations predict singlet $d$-wave
superconductivity \cite{baskaran03,kumar03,kumar04,wang04}
with time-reversal symmetry breaking;
b) charge fluctuations predict a triplet next-nearest
neighbour (NNN) $f$-wave state \cite{tanaka04,motrunich04,foussats05};
c) spin-triplet $f$-wave superconductivity was also proposed based on
phenomenology and symmetry considerations \cite{tanaka03},
due to the topology of the Fermi surface \cite{kuroki04},
from weak-coupling studies of a multiorbital Hubbard model 
\cite{mochizuki05},
and considering spin-orbit coupling \cite{khaliullin04,yanase05a,yanase05b}.

A number of magnetic resonance \cite{fujimoto04,kato06,ihara05}
and $\mu$SR \cite{kanigel04,higemoto04} experiments are consistent with
unconventional triplet superconductivity and exclude time-reversal symmetry
breaking \cite{higemoto04}. Although those results seem at the moment not
conclusive, with some NMR experiments \cite{kobayashi03}
indicating the possibility for singlet $s$-wave superconductivity,
specific heat measurements \cite{yang05} are consistent with the
existence of a superconducting gap with nodal lines.
Based on the results pointing to triplet superconductivity with
nodal lines, $f$-wave symmetry appears as a very promising candidate
(see also \cite{mazin05} for a more detailed discussion).

Besides superconductivity, other features of the electronic
structure point to the proximity of other instabilities.
Photoemission spectroscopy indicate the presence of a pseudogap of
$\sim 20$ meV with a decrease of the density of states at the
Fermi energy as the temperature is lowered \cite{shimojima05}.
Also Raman scattering experiments reported recently a pseudogap
\cite{lemmens05}. Sidebands in the spectra of $E_{1g}$ phonons
suggest that the pseudogap arises from a charge ordering
instability. Such an instability was only observed in
superconducting samples. 
A very recent photoemission experiment \cite{Qian06} was also interpreted 
in terms of the proximity of the system to a charge order phase. 
From the possible mechanisms for
superconductivity discussed above, b) reconciles the possibility
of a charge ordering instability with an unconventional
superconducting state characterized by nodal lines in the order
parameter. However, estimates of $T_c$ for a pure electronic model
lead to extremely low values \cite{motrunich04}.

As a natural extension we have recently considered the interplay
between phonons and electronic correlations \cite{foussats05}.
There, the $t$-$V$ model on the triangular lattice (where $V$ is the
Coulomb repulsion between nearest neighbours) was proposed for the
electronic sub-system. The main effect of $V$ is to bring the
system closer to a $\sqrt{3} \times \sqrt{3}$ charge-density wave 
(CDW) phase where
charge fluctuations are strongly increased. Under these
circumstances, the electron-phonon (e-ph) interaction vertex is
renormalized by charge fluctuations leading to superconductivity
with 
NNN-$f$ unconventional
pairing symmetry
around $x\sim035$.

\begin{figure}\label{Fases}
\begin{center}
\setlength{\unitlength}{1cm}
\includegraphics[width=8cm,angle=0.]{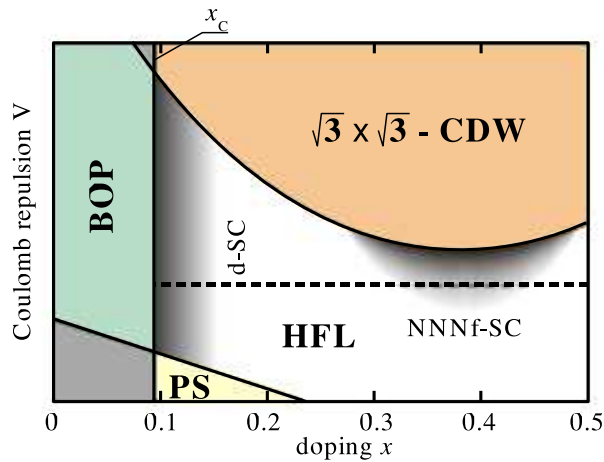}
\end{center}
\caption{(Colour online) Sketch of the obtained generic phase
diagram. The figure shows the phases HFL, BOP, $\sqrt{3}\times
\sqrt{3}$-CDW and PS. Inside the HFL phase superconductivity is
possible in the shading regions. Tendencies to superconductivity
are larger in the regions where the intensity of the shadow is
larger. For instance, making a cut at a given $V$ along the dashed
line we find superconductivity with NNN-$f$ symmetry following a
dome shape around $x\sim 0.38$. 
For low $x$ superconductivity is possible
with $d$-wave symmetry for finite $J$. With decreasing $J$ the
onset of the BOP ($x_c$) moves toward $x=0$ and,
superconductivity, is at the same time, strongly suppressed. The
sketch was presented for $0<x<0.5$ where our theory predicts, in
agreement with experiments, a paramagnetic metal. For $x>0.5$ the
present approach does not reproduce the observed Curie-Weiss
metal. }\label{FaseStutt}
\end{figure}

As mentioned above,
originally, several theories were developed on the basis of the pure 
$t$-$J$ model
\cite{baskaran03,wang04,kumar03}.
Moreover, recent NMR experiments \cite{ning05} point to enhanced low
frequency spin fluctuations before superconductivity sets in. Therefore,
a more general study is needed, where both $V$ as well as the  
antiferromagnetic spin exchange $J$
are considered, in order to clarify the interplay or
competition of electronic instabilities in different parts of the phase 
diagram.

We present here such a study, 
showing that the region where singlet d-wave pairing due to $J$ dominates
is well separated from the one where the NNN f-wave paring appears.

The presence of $J$ triggers a normal state $V$-independent 
instability at low doping.  
This so-called
bond-order-phase (BOP) is found to be mainly of
$d$-wave character and dominated by the exchange interaction $J$.
Phase separation (PS) was also obtained at low doping. 
After identifying the parameter region ($J$-$V$-$x$) where the HFL is stable we have 
studied superconductivity.
As in the case with $J=0$, unconventional
NNN-$f$ pairing near the 
$\sqrt{3}\times \sqrt{3}$-CDW
phase following a dome 
shape around
$x \sim 0.35$, is obtained for expected values of $J$.
The values for $T_c$ obtained in the full model are 
of the order of a few Kelvin like in experiments.
In addition,
for finite $J$  we have found tendencies to
$d$-wave superconductivity 
in the immediate vicinity of the BOP. 
Figure \ref{FaseStutt}
gives a sketch of the phase diagram obtained with our theory.

The paper is organized as follows.
In section II 
the formalism used is summarized, in order to 
allow for a self-contained presentation.
In section
III the stability conditions for the homogeneous Fermi liquid
(HFL) phase are discussed. 
Section IV
is devoted to superconductivity and its possible channels. 
In section V, conclusions and discussions are presented.

\section{ A large-$N$ approach for the $t$-$J$-$V$ model and Feynman rules}

Here we describe our treatment of the $t$-$J$-$V$ model, given by


\begin{equation}\label{Hc}
H = -t \sum_{\langle i,j \rangle,\sigma} (\tilde{c}^\dag_{i\sigma}
\tilde{c}_{j\sigma}+h.c.)
+ J \sum_{\langle i,j \rangle} (\vec{S}_i \vec{S}_j-\frac{1}{4} n_i n_j)
+ V \sum_{\langle i,j \rangle} n_i n_j
\end{equation}

\noindent where $t$, $J$ and $V$ are the  hopping, the exchange interaction
and the Coulomb repulsion, respectively, between nearest-neighbour sites 
denoted by $\langle ij \rangle$.
$\tilde{c}^\dag_{i\sigma}$ and $\tilde{c}_{i\sigma}$ are the
fermionic creation and destruction operators of holes,
respectively, under the constraint that double occupancy is
excluded, and $n_i$ is the corresponding density
operator at site $i$.

Now, we introduce Hubbard operators \cite{hubbard63}
which are related with the fermionic operators as follows
\begin{eqnarray*}
\begin{array}{ll}
  X_i^{\sigma 0}=\tilde{c}^\dag_{i \sigma}\;\;\;\;\; & n_i=(X_i^{\uparrow \uparrow}+
X_i^{\downarrow \downarrow}) \\
  X_i^{0 \sigma}=\tilde{c}_{i \sigma}\;\;\;\;\; & X_i^{\uparrow
  \downarrow}=S_i^+\;\;\;\;\;X_i^{\downarrow \uparrow}=S_i^-  \\
\end{array}
\end{eqnarray*}

The five operators
$X_i^{\sigma \sigma'}$ and
$X_i^{00}$ are boson-like and the four operators
$X_i^{\sigma 0}$ and
$X_i^{0 \sigma}$ are fermion-like.
The names fermion-like and boson-like come from the fact that Hubbard
operators do not verify the usual fermionic and bosonic commutations
rules.

In previous papers \cite{foussats02,merino03,foussats04,greco05} we have
developed a large-$N$ expansion for the $t$-$J$-$V$ model in the
framework of the path integral representation for the Hubbard $X$
operators. In this approach the $X$-operators are treated as
fundamental objects without any decoupling scheme, and hence,
problems that arise in other
treatments are avoided, like considering fluctuations of gauge fields 
that appear in
the slave boson (SB) approach \cite{lee06}.

We start our formalism by extending
the Hamiltonian of the $t$-$J$-$V$ model, to $N$ channels for the spin degrees
of freedom, and rescaling couplings accordingly.
\begin{eqnarray}
H &=& - \frac{t}{N}\sum_{\langle i,j \rangle,p}\;(\hat{X}_{i}^{p 0}
\hat{X}_{j}^{0
p} + h.c.)
+ \frac{J}{2N} \sum_{\langle i,j \rangle;pp'} (\hat{X}_{i}^{p p'}
\hat{X}_{j}^{p' p} - \hat{X}_{i}^{p p} \hat{X}_{j}^{p' p'})
 \nonumber\\
&& + \frac{V}{N}\sum_{\langle i,j \rangle;p p'} \hat{X}_{i}^{p p} \hat{X}_{j}^{p' p'}
-\mu\sum_{i,p}\;\hat{X}_{i}^{p p} \label{eq:H}
\end{eqnarray}

\noindent where the spin indices $\sigma, \bar{\sigma}$ were
extended to new indices $p$ and $p'$ running from $1$ to $N$. In
order to obtain a finite theory in the $N$-infinite limit,  we
rescaled $t$, $J$ and $V$ as
$t/N$, $J/N$ and $V/N$, respectively. In (\ref{eq:H}) $\mu$ is the
chemical potential.

As shown previously, a path integral can be obtained with an
Euclidean Lagrangian $L_E$ given by
\begin{eqnarray}
L_E =  \frac{1}{2}
\sum_{i, p}\frac{({\dot{X_{i}}}^{0 p}\;X_{i}^{p 0}
+ {\dot{X_{i}}}^{p 0}\;
X_{i}^{0 p})} {X_{i}^{0 0}} + H
\end{eqnarray}

\noindent and the following two additional constraints
\begin{eqnarray}
X_{i}^{0 0} + \sum_{p} X_{i}^{p p} - \frac{N}{2}=0 \; \label{eq:v1}
\end{eqnarray}
and
\begin{eqnarray}
\label{ConstraintsWithFermions}
X_{i}^{p p'} - \frac{X_{i}^{p 0} X_{i}^{0p'}}{X_{i}^{0 0}}=
0 \;, \label{eq:v2}
\end{eqnarray}
which are required to satisfy the commutation rules
of $X$-operators. While (\ref{eq:v1}) is the completeness
condition, (\ref{eq:v2}) originates in the above mentioned requirement
to satisfy the commutation rules. For a detailed discussion of the
constraints, and their relation with the commutation rules,
we refer to \cite{foussats02, foussats00}.
In our path integral approach we associate Grassmann and usual
bosonic variables with Fermi-like and boson-like $X$-operators,
respectively.

We now discuss the main steps needed to introduce a large-$N$
expansion \cite{foussats02,foussats04}. First, we
integrate over the boson variables $X^{p p'}$ using
(\ref{eq:v2}). The completeness condition is enforced by exponentiating 
(\ref{eq:v1}) and introducing Lagrange
multipliers $\lambda_i$. We write the boson fields in terms of
static mean-field values, $(r_0, \lambda_0)$ and fluctuation fields
$\delta R_i$, $\delta \lambda_i$, as follows,
\begin{equation}\label{X00Lambda}
X_{i}^{0 0} = N r_{0}(1 + \delta R_{i}) \\
\lambda_{i} = \lambda_{0}+ \delta{\lambda_{i}}, 
\end{equation}
and, we perform the following change of variables for the fermion fields
\begin{equation}\label{fdag}
f^{\dag}_{i p} = \frac{1}{\sqrt{N r_{0}}}X_{i}^{p 0} \\
f_{i p} = \frac{1}{\sqrt{N r_{0}}}\;X_{i}^{0 p}.
\end{equation}

Due to (\ref{ConstraintsWithFermions}), and after using (\ref{fdag}),
the exchange interaction contains four fermion fields that 
can be decoupled in terms of the
bond variable $\Delta_{ij}$ through a Hubbard-Stratonovich transformation,
where  $\Delta_{ij}$ is
the field associated with the quantity
$\sum_{p} f^{\dag}_{j p} f_{i p}/ [(1 + \delta R_{i})
(1 + \delta R_{j})]^{1/2}$. We write the $\Delta_{ij}$ fields in term of
static mean field values and dynamical fluctuations
$\Delta_i^{\eta}=\Delta(1+r_i^\eta+iA_i^\eta)$,  where
 $r_i^{\eta}$ and $A_i^{\eta}$ correspond to the amplitude and the
phase fluctuations of the bond variable,
respectively.
The index $\eta$ takes three values associated with the bond directions
 ${\eta}_{1}=(1,0)$, ${\eta}_{2}=(\frac{1}{2},\frac{\sqrt{3}}{2})$ and
   ${\eta}_{3}=(-\frac{1}{2},\frac{\sqrt{3}}{2})$ of the
triangular lattice.

Introducing the above change of variables
and, after expanding
$1/(1+\delta R)$ in powers of $\delta R$, we arrive at the following
effective Lagrangian:
\begin{eqnarray}\label{Leff}
L_{eff} = -\frac{1}{2}\sum_{i,p}\left(\dot{f_{i p}}f^{\dag}_{i p}
+ \dot{f^{\dag}_{i p}}f_{i p}\right) (1 - \delta R_{i} + \delta R_{i}^{2})
+ t\;r_{0} \sum_{\langle i,j \rangle,p}\;(f^{\dag}_{i p}f_{j p}+h.c.) 
\nonumber \\
- \mu \;\sum_{i,p}\;f^{\dag}_{i p}f_{i p} (1 - \delta R_{i} + \delta R_{i}^{2})
+ N\;r_{0}\;\sum_{i}\delta{\lambda_{i}}\;\delta R_{i}
+ \sum_{i,p} f^{\dag}_{i p}f_{i p}(1 - \delta R_{i}) \; \delta{\lambda_{i}}
\nonumber \\
+ \frac{2 N}{J} \Delta^{2}\sum_{i\eta}\left[({r_{i}^{\eta}})^{2} +
({A_{i}^{\eta}})^{2}\right] 
+ Nr_{0}^{2} (V-  \frac{1}{2}J) \sum_{\langle i,j \rangle} \delta R_{i}\delta R_{j}
\nonumber \\
- \Delta \sum_{\langle i,j \rangle,p} (f^{\dag}_{i p}f_{j p}+f_{jp}^{\dag} f_{ip})
[1 -\frac{1}{2} (\delta R_{i} + \delta
R_{j})+\frac{1}{4}\delta R_{i}\delta R_{j}+\frac{3}{8}(\delta
R_{i}^{2} + \delta R_{j}^{2})]
\nonumber \\
- \Delta \sum_{\langle i,j \rangle,p, \eta} [f^{\dag}_{i p}f_{j p}
({r_{i}^{\eta}}+i {A_{i}^{\eta}})[1 -\frac{1}{2}
(\delta R_{i} + \delta R_{j})]+h.c.],
\end{eqnarray}

\noindent where we have changed $\mu$ to $\mu-\lambda_0$ and dropped
constant and linear terms in the fields.

Looking at the effective Lagrangian (8), the Feynman rules can be
obtained as usual. The bilinear parts give rise to the propagators
and the remaining pieces are represented by vertices. 

To leading order in  $1/N$, we associate with the $N$-component fermion field 
$f_{p}$ a propagator connecting two generic components $p$ and $p'$,
\begin{eqnarray}\label{G0}
G^{(0)}_{pp'}({\bf k}, \nu_{n}) = - \frac{\delta_{pp'}}{\rmi \nu_{n} - E_{k}}
\end{eqnarray}
which is of ${\cal O}(1)$ and where $E_{k}$ is
\begin{eqnarray}\label{Ek}
E_{k} = -2(t r_0+\Delta) (\cos
k_{x}+2\cos\frac{k_{x}}{2}\cos\frac{\sqrt{3}}{2}k_{y} ) - \mu \; ,
\end{eqnarray}
and ${\bf k}$ and $\nu_{n}$ are the momentum and the
fermionic  Matsubara frequency of the fermionic field, respectively.
The fermion variables $f_{ip}$ are proportional to the
$X$-operators (\ref{fdag}) and should not be associated with
the spinons from the SB approach.

\begin{figure}
\vspace{1cm}
\begin{center}
\setlength{\unitlength}{1cm}
\includegraphics[width=8cm,angle=0]{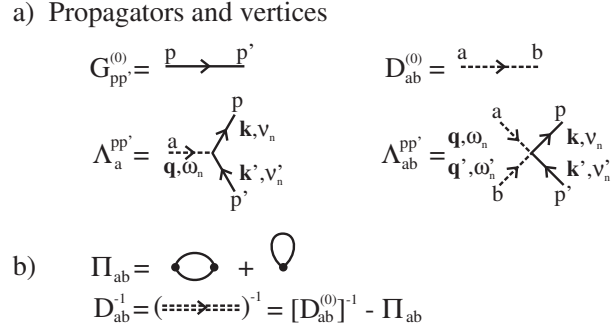}
\end{center}
\caption{Summary of the Feynman rules. a) Solid lines represent the
propagator $G^{(0)}$ (\ref{G0}). Dashed lines
represent the $8 \times 8$ boson propagator
$D^{(0)}$ (\ref{D0})
for the $8$-component field $\delta X^a$. Notice that the component
$(1,1)$ of this propagator is directly associated
with the $X^{00}$ charge operator.
$\Lambda^{pp'}_a$ (\ref{lam3}) and $\Lambda^{pp'}_{ab}$ represent
the interaction between two fermions $f_p$ and one or two bosons
$\delta X^a$ respectively. b) $\Pi_{ab}$ contributions to the irreducible
boson self-energy. Double dashed lines correspond to a dressed boson 
propagators.
}
\label{FR}
\end{figure}

The mean field values $r_0$ and $\Delta$ must be determined by
minimizing the leading order theory. From the completeness
condition (\ref{eq:v1}) $r_0$ is equal to $x/2$ where $x$ is the
electron doping away from half-filling. On the other hand, the expression for
$\Delta$ is
\beq{\label {Delta}}
\Delta = \frac{J}{2N_s} \frac{1}{3}\sum_{k, \eta} \cos(k_\eta) n_F(E_k) \; ,
\eeq
where $n_F$ is the Fermi function and $N_s$ is the
number of sites in the lattice. For a given doping
$x$; $\mu$ and $\Delta$ must be determined self-consistently from
$(1-x)=\frac{2}{N_s} \sum_{k} n_F(E_k)$ and (\ref{Delta}).

We associate with the eight component boson field
\[
\delta X^{a} = (\delta
R\;,\;\delta{\lambda},\; r^{{\eta}_{1}},\;r^{{\eta}_{2}}
,\;r^{{\eta}_{3}},\; A^{{\eta}_{1}},\;
A^{{\eta}_{2}},\;A^{{\eta}_{3}}) ,
\]

\noindent the inverse of the
propagator, connecting two generic components a and b,

\begin{eqnarray}\label{D0}
\fl{D^{-1}_{(0) ab}({\mathbf{q}},\omega_{n})= N \left(
 \begin{array}{cccccccc}
 \gamma_{q} &x/2 & 0&0&0&0 &0& 0 \\
   x/2 & 0 & 0 &0& 0 & 0 &0& 0 \\
   0 & 0 & \frac{4}{J}\Delta^{2} & 0&0 & 0&0 & 0 \\
   0 & 0 & 0 & \frac{4}{J}\Delta^{2} & 0 &0&0& 0 \\
   0 & 0 & 0 & 0& \frac{4}{J}\Delta^{2} &0&0& 0 \\
   0 & 0 & 0 & 0 & 0 & \frac{4}{J}\Delta^{2}&0&0 \\
   0 & 0 & 0 & 0 & 0 &0& \frac{4}{J}\Delta^{2}&0 \\
   0 & 0 & 0 & 0& 0 & 0&0&\frac{4}{J}\Delta^{2} \
 \end{array}
\right),} 
\end{eqnarray}

\noindent where $\gamma_{q}=(4V-2J)(x/2)^{2}\;(\cos
k_{x}+2\cos\frac{k_{x}}{2}\cos\frac{\sqrt{3}}{2}k_{y})$ and the indices $a$, $b$
run from 1 to 8.
${\bf q}$ and $\omega_{n}$ are the momentum and the Bose
Matsubara frequency of the boson field, respectively.

The first component $\delta R$ of the $\delta X^a$ field is
connected with charge fluctuations (\ref{X00Lambda}), i.e. $X^{00}_i=Nr_0(1+\delta R_i)$,
where $X^{00}$ is the Hubbard operator associated with the number of electrons.

The non-quadratic terms in (\ref{Leff}) define three and four leg
vertices:

The three-leg vertex
\begin{eqnarray} \label{lam3}
\Lambda^{pp'}_{a}& =&
(-1)\; \left[\frac{\rmi}{2}(\nu_n + {\nu'}_n)
+ \mu + 2\Delta \sum_{\eta} \cos\left(k_\eta-\frac{q_\eta}{2}\right)\;
\cos\frac{q_\eta}{2},\;1, \right.
\nonumber \\
&& -2\;\Delta\; \cos\left(k_{{\eta}_{1}}-\frac{q_{{\eta}_{1}}}{2}\right)
\; ,-2\;\Delta\; \cos\left(k_{{\eta}_{2}}-\frac{q_{{\eta}_{2}}}{2}\right)
\;,
\nonumber \\
&& -2\;\Delta\;\cos\left(k_{{\eta}_{3}}-\frac{q_{{\eta}_{3}}}{2}\right),
2\; \Delta\; \sin\left(k_{{\eta}_{1}}-\frac{q_{{\eta}_{1}}}{2}\right) \;,
\nonumber \\
&& \left. 2 \Delta\; \sin\left(k_{{\eta}_{2}}-\frac{q_{{\eta}_{2}}}{2}\right) \;,
2 \Delta\; \sin\left(k_{{\eta}_{3}}-\frac{q_{{\eta}_{3}}}{2}\right) \right]\;
\delta^{pp'}
\end{eqnarray}

\noindent
represents the interaction between two fermions and one boson.

The four-leg vertex $\Lambda^{pp'}_{ab}$ represents the interaction between
two fermions and two bosons. The only elements different from zero are:
\[
\fl{\Lambda^{pp'}_{\delta R \delta R}  = \left(\frac{\rmi}{2} (\nu_n + {\nu'}_n)
+ \mu + \Delta\, \sum_{\eta} \cos(k_\eta-\frac{q_\eta+q'_\eta}{2})
\left[ \cos\frac{q_\eta}{2} \, \cos\frac{q'_\eta}{2}\,
+\,\cos\frac{q_\eta+q'_\eta}{2}\right] \right)
\delta^{pp'},
}
\]
\begin{eqnarray}
\Lambda^{pp'}_{\delta R \delta\lambda}=\frac{1}{2}
\;\delta^{pp'} ,\nonumber
\end{eqnarray}
\begin{eqnarray}
\Lambda^{pp'}_{\delta R \; r^{\eta}}= -\Delta \;
\cos(k_\eta-\frac{q_\eta+q'_\eta}{2})\; \cos\frac{q'_\eta}{2}
\;\delta^{pp'}, \nonumber
\end{eqnarray}

\noindent and
\begin{eqnarray}
\Lambda^{pp'}_{\delta R \; A ^{\eta}}= \Delta \;
\sin( k_\eta-\frac{q_\eta+q'_\eta}{2})\; \cos\frac{q'_\eta}{2}
\;\delta^{pp'} . \nonumber
\end{eqnarray}

Each vertex conserves momentum and energy and they are of ${\cal O}(1)$.
In each diagram there is a minus sign for each fermion loop and
a 
symmetry factor.
Feynman rules are given in figure \ref{FR}a.

As usual in a large-$N$ approach,
any  physical quantity can be calculated at a given order just
by counting the powers in $1/N$ of vertices and propagators involved
in the corresponding diagrams.
In the present summary there is no mention of the ghost
fields. They were treated in previous papers \cite{foussats02,foussats04}
and the only role they play is to cancel the infinities given by the two
diagrams of figure \ref{FR}b.

The exchange interaction $J$ enters (\ref{D0}) in two different channels:
a) The term $2J$ in the element (1,1) of $D^{-1}_{(0)}$
is due to the charge-like term,
$ -J/2N \sum_{\langle i,j \rangle;\,pp'} {X}_{i}^{p p} {X}_{j}^{p' p'}$,
of the $t$-$J$-$V$ model.
This term has the same form of the Coulomb term $ V/N
\sum_{\langle i,j \rangle;\, pp'}{X}_{i}^{p p} {X}_{j}^{p' p'}$.
b) The terms $4\Delta^2/J$ in the diagonal of $D^{-1}_{(0)}$ are
due to the exchange-like term,
$ J/2N \sum_{\langle i,j \rangle;\, pp'} {X}_{i}^{p p'} {X}_{j}^{p' p}$,
of the $t$-$J$-$V$ model.

In (\ref{D0}) $V$ is only present in the element $(1,1)$ of $D_{(0)}^{-1}$ and
it is multiplied by $(x/2)^2$ which means that its is
strongly screened, at low doping, by correlations. In addition, the effect
of V is diminished when $J$ is finite.

The bare boson propagator $D_{(0)ab}$ (the inverse of
(\ref{D0})) is ${\cal O}(1/N)$. From the Dyson equation, $D_{ab}^{-1}
= D_{(0) ab}^{-1} - \Pi_{ab}$, the dressed components $D_{ab}$
(double dashed line in figure \ref{FR}b) of the boson propagator can
be found after the evaluation of the $8 \times 8$ boson
self-energy matrix $\Pi_{ab}$. Using the Feynman rules
$\Pi_{ab}$ can be evaluated through the diagrams of figure \ref{FR}b. It 
results
\begin{equation}\label{Piab}
\fl{
\Pi_{ab}({\mathbf{q}},i\omega_{n}) = - \frac{N}{N_{s}}
\sum_{{\mathbf{k}}} h_{a} \; h_{b} \; 
\frac{\left[n_{F}(E_{k+q}) - n_{F}(E_{k})\right]}
{E_{k+q}-E_{k}-\rmi \omega_n}
 - {\frac{N}{N_s}}\;\delta_{a1}\delta_{b1}\;
\sum_{\mathbf{k} }\frac{\varepsilon_{k + q} -
\varepsilon_{k}}{2}\; n_{F}(E_{k}) \; ,
}
\end{equation}

\noindent where
\begin{eqnarray*}
\fl {
h_{a} = \left[ 
\frac{\varepsilon_{k+q}+\varepsilon_{k}}{2},
\; 1,
\; -2 \Delta\;
\cos\left(k_{{\eta}_{1}}+\frac{q_{{\eta}_{1}}}{2}\right), \;
-2 \Delta\; \cos\left(k_{{\eta}_{2}}+\frac{q_{{\eta}_{2}}}{2}\right), \right.
}
\nonumber \\
\fl{
-2 \Delta\; \cos\left(k_{{\eta}_{3}}+\frac{q_{{\eta}_{3}}}{2}\right),
2 \Delta\; \sin\left(k_{{\eta}_{1}}+\frac{q_{{\eta}_{1}}}{2}\right),
\; 2 \Delta\; \sin\left(k_{{\eta}_{2}}+\frac{q_{{\eta}_{2}}}{2}\right),
\; \left.
2 \Delta\; \sin\left(k_{{\eta}_{3}}+\frac{q_{{\eta}_{3}}}{2}\right) 
\right] ,
}
\end{eqnarray*}

\noindent and $\varepsilon_k = -2t\; (x/2) \sum_{\eta} \cos(k_{\eta})$.

The component $(1,1)$ of the dressed boson propagator 
$D_{ab}$ (also called $D_{RR}$)
is related with the charge-charge correlation function $\chi^c_{ij}$.
It can be written as \cite{foussats02, gehlhoff95}
\begin{eqnarray}\label{chi}
\chi^c_{ij}(\tau)=\frac{1}{N} \sum_{pq} \langle T_\tau X_i^{pp}(\tau)
X_j^{qq}(0) \rangle \; .
\end{eqnarray}
and the completeness condition and the relation between
$X^{00}_i$ and $\delta R_i$,
in Fourier space,
\begin{eqnarray}\label{chiv1}
\chi^{c}({\bf{q}},\omega)=-N \left(\frac{x}{2}\right)^2
D_{RR}({\bf{q}},\omega) \; .
\end{eqnarray}
In \cite{foussats02,foussats04,gehlhoff95}
it was pointed out that in ${\cal O}(1)$ the charge-charge
correlation function shows the presence of collective peaks
above the particle-hole continuum.

Finally, one remark is in order at this point. From the completeness
condition (\ref{eq:v1}) we can see that the charge operator
$X^{00}$ is of ${\cal O}(N)$, while the operators $X^{pp}$ are of 
${\cal O}(1)$. This
fact will have the physical consequence that the $1/N$ approach
weakens the effective spin interactions compared to the one
related to the charge degrees of freedom.
Another consequence of this result is the absence, in ${\cal O}(1)$,
of collective excitations (like magnons) in the spin
susceptibility.
The spin-spin correlation function is then
a Pauli like electronic bubble with renormalized band due to
correlations \cite{foussats02, gehlhoff95}.
While there are collective effects in the charge sector in 
${\cal O}(1)$,
those appear in ${\cal O}(1/N)$ in the spin sector.
However, while focusing on superconductivity, that 
in cobaltates occurs at relatively large doping where
the system behaves as a paramagnetic metal,
this fact is not relevant.

\section{Phase diagram: Instabilities of the homogeneous Fermi liquid
}
Before considering the possible electronic instabilities, we recall 
that in leading order we have free fermions with an electronic band $E_{k}$
(\ref{Ek}), renormalized by Coulomb interactions.
From this electronic dispersion we obtain a large Fermi surface (FS)
enclosing the $\Gamma$ point.
First principles calculations \cite{singh03} predict, apart from this FS,
the existence of small pockets near $K$ points.
However, it is important to notice that recent ARPES
experiments \cite{hasan04} do not show the presence of pockets.
Invoking electronic correlations,  a theoretical
explanation for the absence of pockets was given using LDA$+$U
\cite{zhang04} and a strong coupling mean field approach \cite{zhou05}.
These results give an additional support for
considering cobaltates as strongly correlated systems.

Instabilities of the ${\cal O}(1)$ HFL phase
are studied by analyzing the zeros
of ${\rm{Det}} \; D_{ab}^{-1} = {\rm{Det}} \;
(D_{(0) ab}^{-1} - \Pi_{ab})$.
Expanding the determinant by minors along the first row,
it can be written 
in the static limit 
$(\omega_n=0)$ as follows,
\beq \label{Det}
{\rm{Det}} \; D_{ab}^{-1} = f(x, J, {\bf{q}}) \; V
+ g(x, J, {\bf{q}}) \;
\eeq
where $f$ and $g$ are long algebraic expressions
and they were computed numerically in order to study the
instabilities. The fact that the
$\Pi_{ab}$'s are $V$ independent functions was used
in (\ref{Det}).

The system presents two kind of instabilities:
\begin{itemize}
\item[{\it a})] $V$-dependent instabilities: They occur when $f\neq 0$.
From (\ref{Det}) these instabilities take place for given $x$,
$J$ and $\bf{q}$, when the Coulomb potential is $V = V_c=-g/f$.

\item[{\it b})] $V$-independent instabilities: They occur
when the two functions $f$ and $g$
are zero simultaneously.
\end{itemize}

Considering the bare hopping 
$t\sim150$ meV \cite{singh00} and $U_{dd} \sim 3.0-5.5$ eV \cite{chainani04} we obtain, in agreement with other
estimates \cite{yang04,wang04} ($J=4t^2/U$) $J \sim 0.1-0.2$ in units of $t$,
which in the following is considered to be $1$. In what follows we
choose $J=0.2$ which can be seen as an upper bound value.

\subsection{ Charge density wave instabilities}

In this section the stability of the HFL phase will be studied
as a function of $V$, since it corresponds to type {\it a}) above, i.e.\
to the $V$-dependent kind. Two instabilities are found with
critical values $V_{c1}$ and $V_{c2}$.
In figure \ref{Fases} we show the phase diagram in the $V_c-x$ plane
for $J=0.2$.
Regions corresponding to HFL, $\sqrt{3}\times\sqrt{3}$-CDW
($V > V_{c1}$) and phase
separation (PS) ($V < V_{c2}$) are identified.

\begin{figure}
\begin{center}
\setlength{\unitlength}{1cm}
\includegraphics[width=8cm,angle=0.]{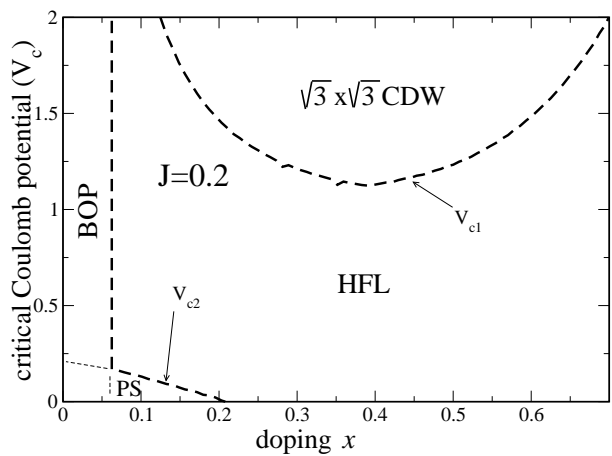}
\end{center}
\caption{
$V$-dependent phase diagram for $J=0.2$.
$V_{c1}$ marks de border between the HFL and the $\sqrt{3} \times \sqrt{3}$-CDW.
$V_{c2}$ marks de border between the HFL and PS.
The vertical dashed line separates the HFL from the BOP
(see subsection {\it 3.2.}).
}
 \label{Fases}
\end{figure}

As usual, PS takes place at ${\bf{q}}=0$.
For a given $J$, the system shows PS for low $x$
and $V$
below the corresponding line marked with $V_{c2}$.
The PS region increases with increasing $J$.

When $V>V_{c1}$ the system enters in a CDW state.
The divergence at the onset of
the CDW is at ${\bf{q}}={\bf{Q}}=(4/3\pi,0)$ corresponding
to a $\sqrt{3} \times \sqrt{3}$-CDW
\cite{motrunich04,foussats05}.
The $V_c-x$ line separating
the HFL from the $\sqrt{3} \times \sqrt{3}$ CDW
phase has a parabola-like shape with a minimum closer to
the doping $x \sim 0.35$ where superconductivity
takes place, with maximum $T_c$,
in cobaltates.

The critical Coulomb repulsion $V_{c1}$
increases with increasing $J$.
As can be seen from the element $(1,1)$ of $D_{(0) ab}^{-1}$ 
in (\ref{D0}), the effect of $V$
is diminished by the presence of  $J$.
For instance, for $x=1/3$,
$V_{c1}$ is $1$ and $1.13$
for $J=0.$ and $J=0.2$,
respectively.
For $J=0$ our phase diagram agrees with the
obtained one in \cite{motrunich04} (see figure 2 of that paper).
The eigenvector corresponding to the zero eigenvalue
of $D^{-1}_{ab}$
is mainly of the form
$(1,0,0,0,0,0,0,0)$ and therefore the instability is concentrated in the
charge sector.

The vertical dashed line 
separating the HFL from the BOP will be discussed
in the next subsection.

\subsection{Bond-order phase}
\begin{figure}
\begin{center}
\setlength{\unitlength}{1cm}
\includegraphics[width=7.8cm,angle=0.]{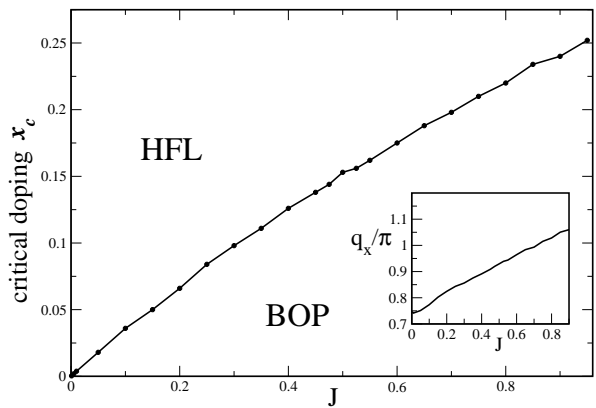}
\end{center}
\caption{
$V$-independent instability. For a given $J$ the HFL is unstable for $x<x_c$
against the BOP. Inset: $q_x$ vs $J$.
}
\label{xcvsJ}
\end{figure}

This instability corresponds to the $V$-independent kind. In figure
\ref{xcvsJ} we show the phase diagram of the model considering
only this instability. For a given $J$, below a critical doping
$x_c$, the HFL is unstable. As can be seen, the instability is strongly
$J$ dependent. When $J \rightarrow 0$, $x_c\rightarrow 0$ showing
that $J$ is the main source for the instability. For instance, for
$J=0.2$, $x_c \sim 0.066$ which corresponds to the vertical dashed
line in figure \ref{Fases}. In the inset of figure \ref{xcvsJ} we show
the $J$ dependence of the critical momentum ${\bf{q_c}}$
where the instability takes place. It is of the form
${\bf{q_c}}=(q_x,0)$ and leads to an inconmensurate instability.

The dominant symmetry involved in the instability is given by the eigenvector 
corresponding to the zero eigenvalue, that in this case 
has the form $\sim (0,0,0,-\sqrt{2}/2,\sqrt{2}/2,0,0,0)$, and hence, it
is in a different sector than the $V$-dependent instabilities. It
means that, for a given $J$,  the amplitude variables
$r^{{\eta}_2}$ and $r^{{\eta}_3}$ of $\Delta^{\eta}$ (the fourth and fifth 
components of $\delta X^a$) are frozen at
${\bf{q}}_c$ for $ x < x_c$. The term $ -\Delta \sum_{\langle i,j \rangle,p,
\eta} (f^{\dag}_{i p}f_{j p}+f^{\dag}_{j p}f_{i p}) r_{i}^{\eta}$,
in the fourth line of our effective Lagrangian (\ref{Leff}) (from
which the components 3, 4 and 5 of the vertex $\Lambda_a^{pp'}$
(\ref{lam3}) are obtained) is, in $q$-space, of the form $2\Delta
\sum_{kqp,\eta} \cos(k_\eta-q_\eta/2) r_q^{\eta} f\dag_{k+q,p}
f_{k,p}$. When the variables $r^{\eta_2}$ and $r^{\eta_3}$ are
frozen, a new hopping-like term 
of the form
$[\cos(q_{cx}/4)d_{xy}({k})+ \sin(q_{cx}/4)p_{x}({k})]$
is generated in the Hamiltonian. 
$p_x$ and $d_{xy}$ are harmonics  of the
triangular lattice \cite{motrunich04}. 
Hence, the instability can be purely of $d$ or $p$ character
depending on whether it takes place at ${\bf q_c}=(0,0)$ or ${\bf
q_{c}}=(2\pi,0)$ respectively.
In-between, both symmetries are mixed. For instance,
for $J=0.2$ we have ${\bf q_c} \sim (0.8,0)\pi$ which means that the
instability is $\sim 65 \%$ $d$ and $\sim 35 \%$ $p$.
Therefore, it is mainly of $d$-wave character at the onset of the instability. 
Since in the triangular lattice $d$-wave symmetry is two-fold
degenerate, the new phase can be a combination of both.

In the case of the $t$-$J$ model on the square lattice, these kind of purely 
electronic  
instabilities were studied using SB \cite{morse90}, Bayn-Kadanoff funtional 
theory \cite{cappelluti99} and path integral large-N approach \cite{foussats04}.
In this case two regimes were obtained: a) For
$J<0.5$, at low doping the system shows an instability whose 
eigenvector is mainly confined to the sector corresponding to phase fluctuations 
$A_i^\eta$ of the bond variables. Therefore, the new phase has 
a complex order parameter
and corresponds to the well known flux phase (FP).
b) For $J >0.5$, at low doping the eigenvector of the instability is confined 
to the sector corresponding to the amplitude variables $r_i^\eta$, and hence,  
corresponds to the BOP. 
This is in fact the order we found in the triangular lattice. 
From the discussion above, and the fact that phase separation is also found 
on the triangular lattice, it seems that instabilities expected on the 
square lattice at high values of $J$ already appear at low values on the
triangular lattice, a feature that may be due to the larger coordination 
number.
We would like to point out that to our knowledge, this kind of analysis has 
not been done before for the $t$-$J$ model on the triangular lattice.

Finally, a word of caution about the BOP is in place here. Our figure
\ref{xcvsJ} shows that at zero doping, the BOP would set in for $J$ finite. 
However, numerical results \cite{bernu92,capriotti99} indicate that the 
antiferromagnetic Heisenberg model on a triangular lattice displays long-range 
N\'eel order for spin $S=1/2$.
The same discrepancy appears on the square lattice with the techniques
metioned above. While due to its bipartite nature, the antiferromagnetic 
order is expected to be rather robust on a square lattice, geometric frustration
on the triangular lattice should render this state much more fragile, such that 
upon doping, RVB-like scenarios \cite{anderson87} appear even more probable in the
present case. Therefore, since
our large-$N$ approach shows phases like the FP and BOP at low doping on a square 
lattice \cite{foussats04}, as other mean 
field aproaches (see above), that are considered as serious 
candidates for underdoped cuprates, we expect the results on the triagular lattice
to be even more trustworthy.

\section{Superconductivity}
Having studied the
stability conditions for the HFL
under the influence of $V$ and $J$, we consider in this section
possible superconducting states.

In \cite{foussats05}
we have proposed that the interplay between
electronic correlations and e-ph interaction is relevant for
describing superconductivity in cobaltates.
To this aim we consider the additional electron-phonon
Hamiltonian $H_{ph}+H_{e-ph}$ where
\begin{equation}\label{Lfon}
H_{ph}=\sum_{i}\omega_{E}\left(a_{i}^{\dag}a_{i}
+\frac{1}{2}\right) \; ,
\end{equation}
and
\begin{equation}\label{Lfon1}
H_{e-ph}=
\frac{g}{\sqrt{N}}
\sum_{i,p}\left(a_{i}^{\dag}+a_{i}\right)X^{pp}_{i}\; .
\end{equation}
In order to obtain a finite theory in the $N \rightarrow \infty$ we
have rescaled the e-ph coupling $g$ to $g/\sqrt{N}$.

From $H_{ph}$ (\ref{Lfon}) we have a free phonon
propagator
\begin{equation}\label{D0ph}
D^{ph}_{0}=\frac{-2\omega_{E}}{\omega^{2}_{n}+\omega^{2}_{E}} \; ,
\end{equation}
which is of ${\cal O}(1)$.
Using the constraint (\ref{eq:v2}), the expression (\ref{Lfon1})
for $H_{e-ph}$ reads
\begin{eqnarray}
  H_{e-ph} =
   \frac{g}{\sqrt{N}}\sum_{ip}(a_{i}^{\dag}+a_{i})
    \frac{{f^{\dag}}_{ip}f_{ip}}    {(1+{\delta} R_{i})} \; ,
\end{eqnarray}
where the relations between $X^{0p}$ and $f_p$ and between $X^{00}$ and
$\delta R$ were also used.
Up to ${\cal O}(1/N)$ only the first term in the expansion of 
$1/(1+\delta R)$ is necessary, leading to
\begin{eqnarray}
  H_{e-ph} =
   \frac{g}{\sqrt{N}}\sum_{ip}(a_{i}^{\dag}+a_{i})(1-{\delta}R_{i})
   f^{\dag}_{ip} {f_{ip}} \; . \label{Lfon2}
\end{eqnarray}

\begin{figure}
\begin{center}
\setlength{\unitlength}{1cm}
\includegraphics[width=6cm,angle=0.]{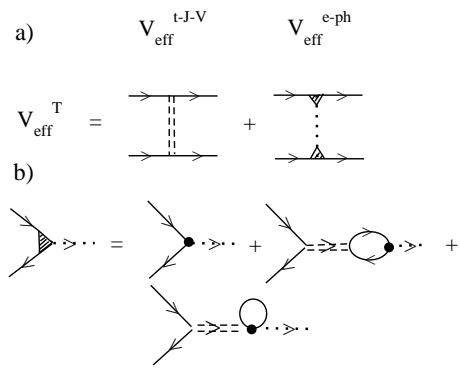}
\end{center}
\caption{a) Total effective paring $V_{eff}^T$ as the sum of a
pure electronic mediated $V_{eff}^{t-J-V}$ and a phonon mediated
$V_{eff}^{e-ph}$ interactions. Solid, double dashed, and dotted
lines are the propagators for fermions, $D_{ab}$, 
and phonons respectively. In $V_{eff}^{e-ph}$, the
bare e-ph vertex, g (solid circle), is renormalized by electronic
correlations as showed in b). The last diagram contains a four leg
vertex proportional to $g$ which is generated when our X-operator
approach is applied (see (\ref{Lfon2})).
 }
 \label{Vertph}
\end{figure}

Next, we discuss the effective interaction between carriers. 
Fluctuations in ${\cal O}(1/N)$ give rise to interactions among fermions. In
${\cal O}(1/N)$ there are two contributions to the total pairing
effective potential $V_{eff}^T$ shown in figure \ref{Vertph}a. The
first diagram of figure \ref{Vertph}a shows the pairing effective
potential, $V_{eff}^{t-J-V}$, for the pure $t$-$J$-$V$ model while the
second diagram shows the e-ph pairing potential $V_{eff}^{e-ph}$.
Notice that $1/N$ self-energy corrections in the fermionic propagator
(\ref{G0}) are not necessary for calculating $V_{eff}$ in ${\cal O}(1/N)$.

Due to the rescaling of the e-ph interaction, $g$,
superconductivity from phonons also appears in ${\cal O}(1/N)$ and therefore,
it can be treated on an equal footing to superconductivity in
the pure $t$-$J$-$V$ model.

The new contribution to the e-ph pairing potential is the vertex
(dark triangle) in the second diagram of figure \ref{Vertph}a which
represents the e-ph interaction renormalized by electronic
correlations. The diagram of figure \ref{Vertph}b shows that the
renormalization of the bare vertex $g$ is due to the electronic
correlations of the pure $t$-$J$-$V$ model, which will be the main
contribution to our results.
Notice that the e-ph vertex is not renormalized by the e-ph
interaction. Hence, we assume, as in usual metals, that Migdal
theorem is valid.

Using our Feynman rules the renormalized e-ph vertex $\gamma$ is

\begin{eqnarray}\label{gammJ}
\fl{
    \gamma(q,k',k'',\omega_{n},\nu'_{n},\nu''_{n})} &=&
    gN\left(\frac{x}{2}+2\Delta\frac{1}{N_{s}}\sum_{k,\eta}
    \cos\frac{q_{\eta}}{2}\cos\left(k+\frac{q}{2}\right)_{\eta}\frac{[\;n_{F}(E_{k + q} ) -
n_{F}(E_{k} )]} { E_{k + q} - E_{k} -\rmi \omega_n} \right)
\nonumber \\
&&\times D^{R b}(q,\omega_{n})  \Lambda_{b}^{pp} \; ,
\end{eqnarray}
where $\Lambda_b^{pp}$ is given by (\ref{lam3}).

In $J=0$ case, the renormalized e-ph vertex can be written as
\begin{eqnarray}\label{gammJ0}
\fl {\gamma(q,k',k'',\omega_{n},\nu'_{n},\nu''_{n})} &=&
Ng\frac{x}{2}\left\{\left[{\frac{\rmi}{2}}(\nu'_{n}+\nu''_{n})
+ \mu\right]D_{RR}(q,\omega_n)
+ D_{\lambda R}(q,\omega_n)\right\} \; ,
\end{eqnarray}
where the relevant contribution comes from the charge-charge correlation
$D_{RR}$ (\ref{chiv1}).
Vertex corrections obtained by us
are similar to the early calculation of \cite{kulic94} which were used before for studying
transport \cite{zeyher96} and isotope effect in cuprates \cite{greco99}.

In weak coupling we can evaluate
the effective interactions on the FS, i.e.\ for
$\omega_{n}=\nu'_{n}=\nu''_{n}=0$ and the momentum $\bf{q=k-k'}$ with
$\bf{k}$ y $\bf{k'}$ on the FS. Then,
the total pairing effective potential is
\begin{equation}\label{VT}
    V^{T}_{eff}({\mathbf{k},\mathbf{k'}})=
    V^{t-J-V}_{eff}({\mathbf{k},\mathbf{k'}})+
    V^{e-ph}_{eff}({\mathbf{k},\mathbf{k'}})\; ,
\end{equation}
where, using the Feynman rules,
\begin{equation}
V^{t-J-V}_{eff}(\mathbf{k},\mathbf{k'})=\Lambda_{a}^{pp}
D^{ab}(\mathbf{k}-\mathbf{k'}) \Lambda_{b}^{pp} \; ,
\end{equation}
and
\begin{equation}\label{Veffph}
V^{e-ph}_{eff} (\mathbf{k},\mathbf{k'})= -\frac{\lambda}{N(0)}
\left[
\gamma^{*} (\mathbf{k},\mathbf{k'}) \right]^{2} \; ,
\end{equation}
where $\lambda=\frac{2 g^2}{\omega_E} N(0)$ is the bare
dimensionless e-ph coupling and $N(0)$ is the bare electronic
density of states. In (\ref{Veffph}) $\gamma^*=\gamma/g$.

In the following we choose
$\lambda=0.4$.
To our knowledge the value of $\lambda$ is not
known for cobaltates yet. However, recent experiments suggest that it is
nonnegligeable \cite{lupi04}. $\lambda=0.4$ is of the order of recent 
estimates \cite{yada05,rueff06}.

Without correlations, $\gamma^*=1$ and $V_{eff}^{e-ph}$ is
the usual pairing potential used in BCS theory which
in conventional metals leads
to superconductivity in the isotropic $s$-wave channel. However,
in such a case, the characteristic dome shape observed in cobaltates would 
not be expected, because there is no reason for a
strongly doping dependent bare e-ph coupling $\lambda$.
Recently, Yada and Kontani \cite{yada05}, using a $d-p$ model for
$NaCoO_2$, found evidence for phonon mediated superconductivity in the
s-wave channel, so that, as they pointed out, other effects, such as 
electronic correlations, are
necessary in order to stabilize an anisotropic pairing.

We use the  effective potentials to compute the dimensionless
effective couplings in the different pairing channels or
irreducible representations of the order parameter.
The critical temperatures $T_c$, can be then estimated
from: $T_{ci}=1.13 \omega_c \exp(1/\lambda_i)$, where $\omega_c$
is a suitable cutoff frequency.

The effective couplings $\lambda_i$ with
different symmetries are defined as \cite{merino03}:

\begin{eqnarray}
\lambda_i=\frac{1}{(2 \pi)^2}
\frac{\int (d {\bf k} /|v_{\bf k}|) \int (d {\bf k'}/|v_{\bf k'}|)
g_i({\bf k'})
V_{eff}({\bf k',k}) g_i({\bf k})}{
\int (d {\bf k}/|v_{\bf k}|)  g_i({\bf k})^2 }
\end{eqnarray}

\noindent where the functions $g_i({\bf k})$, encode the different pairing
symmetries (see Table I of \cite{motrunich04} for the
triangular lattice), and $v_{\bf k}$ are the quasiparticle velocities at
the Fermi surface. The integrations are restricted to the Fermi
surface. $\lambda_i$ measures the strength of the interaction
between electrons at the Fermi surface in a given symmetry channel
$i$. If $\lambda_i > 0$, electrons are repelled. Hence,
superconductivity is only possible when $\lambda_i <0$.

\begin{figure}
\begin{center}
\setlength{\unitlength}{1cm}
\includegraphics[width=9.5cm,angle=0.]{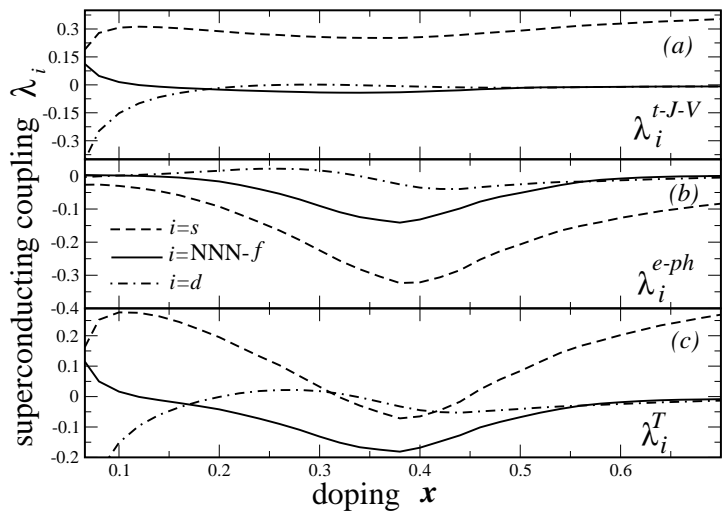}
\end{center}
\caption{
Superconducting couplings $\lambda_i$
 in the $s$ (dashed line), $d$ (dotted dashed line) and NNN-$f$ (solid line)
channels for a) the pure $t$-$J$-$V$ model, b) the e-ph model and c)
the total case. Results are for $J=0.2$,  and $V=0.9 V_c$  
which locates the system near the $\sqrt{3} \times
\sqrt{3}$-CDW phase. The bare e-ph superconducting coupling was
chosen to be $\lambda=0.4$. } \label{lamS}
\end{figure}

Figure \ref{lamS} shows results for
the most relevant symmetry channels as $s$, $d$ and NNN-$f$ for
$J=0.2$ and $V \sim 1$ close to the $\sqrt{3} \times \sqrt{3}$-CDW 
(see figure \ref{Fases}).
The curves were cut at $x \sim 0.066$ where the
BOP instability takes place.
Figure \ref{lamS}a corresponds to the pure $t$-$J$-$V$ model. As
expected $\lambda_s^{t-J-V} > 0$ (dashed line) hence, electrons
are repelled indicating no tendencies to superconductivity with
$s$-symmetry.

In the NNN-$f$ channel (solid line) we have obtained
small negative values for $\lambda_{NNN-f}^{t-J-V}$ with a shallow minimum 
around
$x \sim 0.38$ suggesting the possibility for superconductivity.
However, in the most favorable case
$\lambda_{NNN-f}^{t-J-V} \sim -0.04$ implying that an
unrealistic cut-off $\omega_c$ is necessary in order to obtain a
value of a few Kelvin for $T_c$.
This feature remains valid even closer to the
$\sqrt{3} \times \sqrt{3}$-CDW phase.

The dotted-dashed line shows results for
$\lambda_d^{t-J-V}$. 
Negative superconducting couplings are found 
at small doping, $x<0.15$,  where
$\lambda_d^{t-J-V}$ can take robust values $\sim -0.3$,
indicating that 
superconductivity in the d-wave channel may be expected at such low doping.
Our calculations show that $\lambda_d^{t-J-V}$ is independent of $V$.
On the other hand, $\lambda_d^{t-J-V}$ is
strongly $J$ dependent, vanishing fast when 
$J \rightarrow 0$.
Comparing figures \ref{Fases} and \ref{lamS} it can be seen that
$d$-wave superconductivity occurs near the onset of BOP which is also 
of $d$-wave character and triggered by $J$.
For the
doping range where superconductivity takes place in cobaltates, $x \sim 0.35$,
there is no indication for $d$-wave pairing.

Figure \ref{lamS}b shows results for the e-ph case. 
$\lambda_s^{e-ph}$ (dashed line) would suggest that superconductivity 
could be expected in the s-wave channel
around $x\sim 0.38$ following a dome shape as in cobaltates.
However, in a
Gutzwiller description the s-wave order parameter would be exactly
zero. 
In contrast to that the enforcement of the large-$N$ non-double
occupancy constraint (\ref{eq:v1}), namely, that only $N/2$ out of
the total $N$ states at a given site can be occupied at the same
time, makes s-wave superconductivity possible 
in spite of strong correlations  
(see \cite{greco99} for discussions).

$\lambda_{NNN-f}^{e-ph}$ (solid line) shows
robust
negative superconducting couplings following a dome
shape around $x \sim 0.38$. Notice that
$\lambda_{NNN-f}^{e-ph}\sim -0.16$
which is
a factor 4 larger than for the pure $t$-$J$-$V$ model.
NNN-$f$ superconductivity is not very sensitive to the
value of $J$ and the only requirement is that the system must
be close to a charge instability (in this case a 
$\sqrt{3} \times \sqrt{3}$-CDW).

The
dotted-dashed line in figure \ref{lamS}b shows the projection of the e-ph coupling on the 
$d$-wave channel,
$\lambda_d^{e-ph}$. In the uncorrelated case 
$\lambda_d^{e-ph}$ is exactly zero, but electronic correlations are 
responsible 
for the weak modulation with doping of  
$\lambda_d^{e-ph}$.
For low doping, 
$\lambda_d^{e-ph}$ can take small but negative values of the order of 
$\lambda_d^{e-ph} \sim -0.02$ (not well appreciated in the scale of figure \ref{lamS}).
These small and negative values for  
$\lambda_d^{e-ph}$ increse with $J$. An appreciable strength of
$\lambda_d^{e-ph}$ can only be obtained 
for unrealistic values of $J$. As we mentioned above the renormalized 
e-ph vertex 
is mainly dominated by charge fluctuations and then, NNN-$f$ pairing proves  
correlation effects better than $d$-wave pairing. 

Figure \ref{lamS}c shows results for the total coupling
$\lambda_i^{T}=\lambda_i^{t-J-V} + \lambda_i^{e-ph}$.
The solid line
($\lambda_{NNN-f}^{T}$) shows the strongest tendencies to
superconductivity. The dome shape around $x\sim0.38$ is clearly
present. While $\lambda_{NNN-f}^{t-J-V}$ is small but negative,
$\lambda_{NNN-f}^{T}$ increases slightly with respect to
$\lambda_{NNN-f}^{e-ph}$.

The situation is different for the $s$-wave channel;
as
$\lambda_s^{t-J-V} > 0$, $\lambda_s^T$ will be smaller than
$\lambda_s^{e-ph}$. For instance,
$\lambda_s^T \sim
-0.05$ at $x \sim 0.38$, as a result of strong correlations.

Finally, as the contribution $\lambda_d^{e-ph}$ (dotted dashed
line in panel b) is  very small, the total effective coupling
$\lambda_d^T$ is close to $\lambda_d^{t-J-V}$. In the triangular
lattice the $d$-wave channel is degenerate. Our calculation
determines the leading symmetry of the superconducting order
parameter but not its value. However, our results are consistent
with previous mean field studies \cite{kumar03, wang04} where
superconductivity was found with $d_1+\rmi d_2$ symmetry. We would like
to remark that these mean-field studies assume the pure
$J$ term as the effective interaction. In our calculation we have
included fluctuations through the infinite series of bubbles in
the evaluation of the propagator $D_{ab}$.

We conclude that the $t$-$J$-$V$ model alone 
would support superconductivity in the d-wave channel
at low doping ($x<0.15$ for $J=0.2$). To our knowledge there
are at present no data for such doping levels concerning 
superconductivity.
On the other hand, while e-ph interaction essentially introduces 
no changes in the d-wave channel, it effectively couples to
charge fluctuations close to a charge ordering instability.
Near
charge-order, charge-fluctuations renormalize the e-ph effective
interaction in such a way that superconductivity with triplet 
NNN-$f$ symmetry is favored around $x\sim 0.38$.

\begin{figure}
\begin{center}
\setlength{\unitlength}{1cm}
\includegraphics[width=8cm,angle=0.]{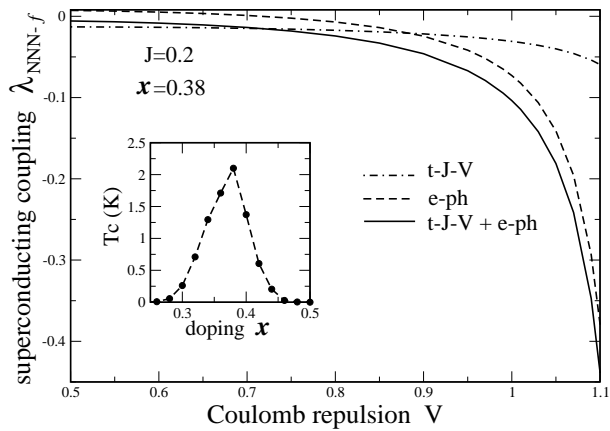}
\end{center}
\caption{ Superconducting coupling in the NNN-$f$-channel as a
function of $V$ approaching the critical value $V_c$ for $J=0.2$.
The figure is presented for $x=0.38$ where the largest
superconducting coupling was obtained. Results for the pure
$t$-$J$-$V$ model (dotted-dashed line), pure e-ph model (dashed line)
and the total case (solid line) are presented showing that, when
phonons are included, a large window near the  $\sqrt{3}
\times \sqrt{3}$-CDW phase exists where the system presents 
indications for superconductivity. Inset: $T_c$ vs doping for the
NNN-$f$ channel. The BCS formula was used for the estimation of
$T_c$ (see text).
 }\label{Tcvsdp}
\end{figure}

In order to see the influence of the proximity to the
charge order on superconductivity, 
in figure \ref{Tcvsdp} we show, for $x=0.38$, the
values of $\lambda_{NNN-f}^{t-J-V}$ (dotted-dashed line),
$\lambda_{NNN-f}^{e-ph}$ (dashed line) and $\lambda_{NNN-f}^{T}$
(solid line) as a function of $V$. 
For the pure $t$-$J$-$V$ model
the values of $\lambda_{NNN-f}^{t-J-V}$
are very small even very close to the 
$\sqrt{3} \times \sqrt{3}$-CDW. 
In contrast, when e-ph interaction is included,
there is a large window for the parameters for which
superconductivity may be possible. As $\lambda_{NNN-f}^T$
increases with $V$, $T_c$ will also increase. It is known from
experiments that $T_c$ increases with water
inclusion \cite{sakurai04} supporting the view that the
increasing of $V$ mimics the increasing of water content. For
small $V$, $\lambda_{NNN-f}^T$ is very small in agreement with the
non-existence of superconductivity in nonhydrated samples.

Although at this point a quantitative comparison with
experiments is beyond the scope of our analysis, it is still
of interest to make a qualitative
estimate of $T_c$.
Since NNN-$f$ superconductivity is in our case
mainly mediated by phonons, we
consider as a relevant cut-off 
corresponding energy scale. 
Recent first principles lattice dynamics calculations
\cite{li04} show the existence of optical phonons as high as $75
$ meV and for simple estimates we consider $\omega_E=40
$ meV \cite{rueff06}. Using the values for $\lambda_{NNN-f}^{T}$
(solid line in panel c of figure \ref{lamS}) we show, in the inset of
figure \ref{Tcvsdp}, results for $T_c$. 
Our crude estimate gives $T_c \sim 2K$ which is of the order of the experimental value 
$T_c \sim 5K$.
A priori, this result may appear trivial because the e-ph interaction is 
certainly efficient for leading to superconductivity 
with $T_c$ in the scale of a few Kelvin in usual metals.
However, there are two features in our results that would be absent when 
correlations are left aside. Firt, the pairing channel is an unconventional 
one, a result that would not be possible considering e-ph interaction alone. 
Second, a dome shape is obtained for $T_c$, again a fact that would be 
missing by considering e-ph interaction alone.

\section{Conclusions and discussions}
We have proposed that the $t$-$J$-$V$  model plus phonons,
on the triangular lattice, may
describe
superconductivity in hydrated cobaltates.

Before studying superconductivity we have presented the phase
diagram of the model. Two types of instabilities were found: a) 
$V$-dependent instabilities. These instabilities are dominated by the
short range Coulomb repulsion $V$.  For $V$ larger than a critical
value $V_c$ the system enters a $\sqrt{3}\times \sqrt{3}$-CDW
phase. PS was also obtained for small $x$ and $V$. b)
$V$-independent instability. This instability is dominated by $J$
and is common to the pure $t$-$J$ model. For a given $J$,  the
system is unstable for doping smaller than a critical one $x_c$.
This new phase occurs at an incommensurate momentum $q_c$ and it
is called BOP. It was found that $x_c \rightarrow 0$ when $J
\rightarrow 0$.

These phases delimit the region of parameters ($x$, $J$ and $V$)
where the HFL is stable. Paring was calculated in this last
region.

Near the $\sqrt{3} \times \sqrt{3}$-CDW  state
the e-ph vertex is renormalized by electronic correlations
developing an anisotropy in $k$-space due to the coupling with
charge-fluctuations.
This anisotropy favors superconductivity  with NNN-$f$ symmetry
when the renormalized vertex is used for calculating phonon mediated pairing.
Besides the possibility of  anomalous pairing, the model
shows possible superconductivity
following a dome shape around $x \sim 0.35$ with values for
$T_c$ of the order of a few Kelvin.
We have found that the above results
are
robust against the value of $J$ around $x \sim 0.35$.

In addition to NNN-$f$ superconductivity for doping $x \sim 0.35$, 
for finite $J$ we have found the possibility of $d$-wave pairing at  low
doping. For instance, for $J=0.2$, on the high side 
for possible values 
of $J$, $d$-wave pairing can be expected for $x<0.15$. 
In contrast to NNN-$f$ pairing, $d$-wave pairing is not affected by $V$ 
suggesting that
superconductivity 
in that channel could exist even without hidratation.
To our knowledge superconductivity was not investigated for samples
at such small doping levels. 

It was recently proposed \cite{marianetti04}, and studied
ex\-perimen\-tally \cite{sakurai04}, that hydration causes the
electronic structure to be more two dimensional.
Notice that in \cite{sakurai04} it was shown that $T_c$ decreases with
decreasing lattice parameter $c$.
We think that due
to this effect the Coulomb repulsion $V$ may be less screened when the
system is hydrated. If $V$ is small when the system is not
hydrated, phonons will favor superconductivity only in the
$s$-wave channel. However, the strong repulsion in this channel
from the $t$-$J$-$V$ model (figure \ref{lamS}a) suppresses pairing from
phonons. This may be the reason for the nonexistence of
superconductivity in non-hydrated cobaltates.

As NNN-$f$ superconductivity has a large contribution from phonons we
expect a large isotope coefficient. Our theory also predicts a
rather constant isotope coefficient along the dome in contrast to
the strong doping dependent isotope coefficient in cuprates
\cite{frank94}. To our knowledge, isotope effect experiments are
still not available for cobaltates.
We expect that the
improvement in single crystals preparation \cite{krockenberger05} will be
useful for isotope experiments in the near future.
We consider this experiment as a strong testing for our approach.

We would like to remark about some analogies
between cobaltates and organic materials. 
As in organic materials \cite{dressel03},
optical conductivity experiments in cobaltates \cite{hwang05} 
show the presence
of low energy features which can be associated with the proximity of the
system to the charge-order.
Merino and McKenzie \cite{merino01} pointed out that the proximity
to the charge-order is relevant for superconductivity with anomalous
paring in organic systems.
As  our
scenario predicts a $\sqrt{3} \times \sqrt{3}$-CDW
state for $V>V_c$ it will be
interesting to see if a further inclusion of water can trigger the
$\sqrt{3} \times \sqrt{3}$-CDW
phase or, at least, if low energy optical features
are reinforced with hydration.

Recent reports discuss the possibility for singlet $s$-wave
superconductivity \cite{kobayashi03}
and the coexistence of
$s$-wave and unconventional pairing
\cite{chen05}. Such a situation could be reached in our
case by increasing the bare $\lambda$ from $\lambda=0.4$, 
since then, the total superconducting couplings
$\lambda_s^T$ and $\lambda^T_{NNN-f}$ become
more attractive and, for $\lambda > 1$ both symmetries are nearly
degenerated. However, lacking detailed information about the
e-ph coupling, we take a cautious value for $\lambda$, that is
already sufficient to trigger superconductivity, and as shown
above, of unconventional type.

\ack{
The authors thank to J Merino, Y Krockenberger
for valuable discussions.
M Bejas thanks to DAAD for financial support and the University of
Stuttgart for hospitality.}

\end{document}